\documentclass[aip,amsmath,amsthm,amssymb,citeautoscript,reprint]{revtex4-2}
\usepackage{graphicx, subfig, xcolor}
\usepackage{physics, siunitx}
\usepackage{booktabs}

\usepackage[todonotes={textsize=tiny}, final]{changes}

\usepackage{hyperref}
\hypersetup{
    pdfauthor = {Amartya Bose},
    pdftitle = {Quantum-Classical Hierarchical Equations of Motion},
    colorlinks = true,
}

\begin{document}
\title{Quantum-Classical Hierarchical Equations of Motion}
\author{Amartya Bose}
\email{amartya.bose@tifr.res.in}
\email{amartya.bose@gmail.com}
\affiliation{Department of Chemical Sciences, Tata Institute of Fundamental Research, Mumbai 400005, India}
\begin{abstract}
  We develop a quantum-classical hierarchical equations of motion
  (QC-HEOM) approach for simulating non-Markovian open quantum
  systems. The method combines the ensemble-averaged classical path
  reference of the quantum-classical path integral formalism with a
  hierarchy of auxiliary quantum influence functionals. By
  incorporating thermal fluctuations through an ensemble average over
  reference trajectories, the hierarchy is required to represent only
  the residual quantum memory associated with the imaginary part of
  the bath response function. Consequently, unlike conventional
  hierarchical equations of motion, QC-HEOM does not require Matsubara
  or Pad\'e expansions of the thermal kernel and exhibits only weak
  temperature dependence of the hierarchy size.  Furthermore, because
  thermal fluctuations are supplied through reference classical
  trajectories, the framework \added{provides a natural route toward extensions}
  beyond harmonic baths and enables the incorporation of anharmonic and
  molecular environments through externally generated trajectories. We derive
  the formalism and demonstrate its exactness for a harmonic bath. Applications
  to an asymmetric spin-boson model and the seven-site Fenna--Matthews--Olson
  complex illustrate the accuracy of QC-HEOM. It reproduces benchmark
  quasi-adiabatic path integral and hierarchical equations of motion results
  while requiring substantially fewer auxiliary objects, particularly at low
  temperatures. These results establish QC-HEOM as an efficient framework for
  treating residual quantum memory in quantum-classical descriptions of
  open-system dynamics. The separation of thermal fluctuations from residual
  quantum memory through the use of Wigner trajectories \added{suggests} an approximate
  route toward hierarchical treatments of complex anharmonic environments beyond harmonic bath simulations.
\end{abstract}

\clearpage % Forces current column content to finish and clear
\onecolumngrid
\listofchanges[title={Major Revisions}, show=comment]
\clearpage % Ensures the list is finished before going back to 2 columns
\twocolumngrid

\maketitle
\section{Introduction}
The numerical challenges of simulating the time-evolution of a quantum
system under the Schr\"odinger equation scales exponentially with
dimensionality. In systems such as exciton- or charge- transfer, it is
often seen that the actual event of interest is actually localized
within a much smaller dimensional Hilbert space with the other degrees
of freedom modulating the dynamics. This realization has lead to a
variety of methods that develop a system-environment decomposition,
where the environment is traced out, leading to a non-Markovian
equation of motion.

Two broad classes of methods exist within the system-environment
decomposed approaches. First, we have the category of numerically
exact methods, which can only be applied to the case of the
environment being represented as harmonic baths. The family of methods
based on the quasi-adiabatic propagator path integral
(QuAPI),\cite{makriTensorPropagatorIterativeI1995,
  makriTensorPropagatorIterativeII1995,
  makriBlipDecompositionPath2014, makriIterativeBlipsummedPath2017,
  strathearnEfficientNonMarkovianQuantum2018,
  bosePairwiseConnectedTensor2022,
  boseMultisiteDecompositionTensor2022} and the methods related to the
hierarchical equations of motion (HEOM)\cite{tanimuraTimeEvolutionQuantum1989,
ishizakiQuantumDynamicsSystem2005, tanimuraStochasticLiouvilleLangevin2006,
shiEfficientHierarchicalLiouville2009, xuTamingQuantumNoise2022,
tanimuraRealtimeImaginarytimeQuantum2015, tanimuraNumericallyExactApproach2020,
moixHybridStochasticHierarchy2013, keExtensionStochasticHierarchy2017}~\comment{Original HEOM reference added.} are two
of the most well-known exact methods in the community. From their inception more
than three decades ago, these two methods have become the standard for benchmark
applications and accuracy. The main shortcoming of these methods has
historically been their extraordinarily high cost and complexity. Several
algorithmic improvements have made their application to simulation of large
systems possible.

The other category comprises of the ubiquitous mixed quantum-classical
methods, that relegate the treatment of the environment to classical
trajectories and evolve the system using approximate quantum mechanics
in response to these trajectories. The earliest mixed
quantum-classical method is probably Ehrenfest dynamics or
self-consistent field dynamics. Approaches based on surface hopping
dynamics\cite{tullyMixedQuantumClassical1998,
  tullyMolecularDynamicsElectronic1990,
  tullyNonadiabaticMolecularDynamics1991,
  tullyPerspectiveNonadiabaticDynamics2012} are also extremely popular
for approximating the non-adiabatic evolution of systems. The
quantum-classical path integral
(QCPI)\cite{lambertQuantumclassicalPathIntegralI2012,
  lambertQuantumclassicalPathIntegralII2012} provides a rigorous and
systematic way of approximating the path integral results using
(quasi-)classical trajectories for anharmonic systems. While
approximate for anharmonic systems, QCPI is exact in the limit of
harmonic baths and reduces to QuAPI.

Under QCPI, the system's reduced density matrix is written as an
environment phase space integral of the so-called ``quantum influence
function,'' which is evaluated as a path integral with free solvent-driven
reference
propagators\cite{banerjeeQuantumClassicalPathIntegral2013}. Makri showed that
\added{these reference propagators, also called the ``ensemble averaged
classical path'' (EACP) reference propagators,}\comment{EACP defined} themselves
account for the ``classical memory'' corresponding to the real part of the bath
response function\cite{makriQuantumclassicalPathIntegral2015}. The path integral
incorporates the ``quantum memory.'' It has also been suggested that restricting
the full anharmonic trajectories for the reference propagator part and
approximating the quantum memory using the harmonic mapping can be a convenient
approximation that avoids exponential proliferation of classical trajectories
while keeping the path integral with analytical influence functional
coefficients\cite{wangQuantumclassicalPathIntegral2019, bosePhaseSpacePath2018}.

Despite these improvements, QCPI still requires path integral simulations for
every Monte Carlo sample of the solvent phase space. This can, depending on the
parameters, be quite computationally intensive. On the other hand, the
application of HEOM to non-adiabatic dynamics, even with all the developments,
is limited to bosonic or fermionic baths. (For vibrational problems, a HEOM-like
quantum hierarchical Fokker-Planck
equations\cite{tanimuraNumericallyExactApproach2020} can be used under Wigner
approximation.) In this work, we present a combination of the two ideas, where,
as in QCPI, the classical memory is accounted for by solvent-driven
references, however the quantum memory is incorporated using HEOM. This avoids
the path integral calculation altogether leading to better scaling.
Additionally, the number of poles required for the HEOM calculation is
significantly reduced, because the real part of the bath correlation function is
completely accounted for by the reference, thus avoiding the proliferation of
poles required to represent the thermal Bose-Einstein distribution in the real
part. \added{Moreover, because this quantum-classical HEOM (QC-HEOM) is based on
classical trajectories, it offers a natural framework for future semiclassical
incorporation of anharmonic effects.}

\added{Stochastic HEOM~\cite{moixHybridStochasticHierarchy2013,
keExtensionStochasticHierarchy2017} provides another approach for treating the
real part of the bath response function by introducing a stochastic unraveling
through a Hubbard–Stratonovich transformation. Recent extensions have enabled
the treatment of anharmonic baths by incorporating higher-order multitime
correlation functions~\cite{hsiehUnifiedStochasticFormulation2018,
hsiehUnifiedStochasticFormulation2018a}. QC-HEOM follows a different philosophy.
Rather than introducing an auxiliary stochastic field, it identifies the
fluctuation contribution associated with the real part of the bath response
function with the solvent-driven reference trajectories already present in the
QCPI framework. The remaining dissipative contribution is treated
hierarchically. The stochastic variables sampled in the two approaches are
therefore fundamentally different, stochastic HEOM samples realizations of a
Gaussian colored-noise field, whereas QC-HEOM samples initial conditions of
physical bath trajectories from a thermal Wigner distribution. This connection
naturally inherits the flexibility of QCPI and permits future extensions toward
anharmonic or atomistic environments within the harmonic backreaction
approximation. A detailed comparison of stochastic sampling strategies and their
convergence properties will be explored in future work.}\comment{Relation with
stochastic HEOM} For harmonic environments, QC-HEOM remains formally equivalent
to the underlying QuAPI while replacing path-integral propagation with a
hierarchical equation-of-motion treatment of the residual quantum memory.

In Section~\ref{sec:method}, \added{we derive QC-HEOM from the fundamental path
integral expressions and discuss how the reference trajectory formulation
provides a possible extension toward anharmonic environments.} Applications
involving asymmetric spin-boson system and the seven-site model for the
Fenna--Matthews--Olson complex are used to compare the performance and
convergence of QC-HEOM to QCPI and HEOM respectively in Sec.~\ref{sec:numerics}.
It is seen that even relatively low depths of hierarchy suffices to give fully
converged results, with the biggest correction happening from the EACP dynamics
to the dynamics corresponding to depth 1. \added{Because the complexity grows
only with the number of terms in an exponential representation of the bath
response function, this proves to be very efficient.} Finally, some concluding
remarks and future directions are presented in Sec.~\ref{sec:conclusion}.

\section{Method}\label{sec:method}
For simplicity of development, let us consider a system coupled with a
bath of harmonic oscillators described by
\begin{align}
  H &= H_0(\hat{s}) + H_\text{SB}(\hat{q}, \hat{s}),
\end{align}
where $H_0$ is the system Hamiltonian and $H_\text{SB}$ is the bath
Hamiltonian (including the system-bath interaction terms). Using path
integrals, one can express the time-evolved reduced density matrix
corresponding to the system starting from a separable initial
condition,
$\rho(0)=\tilde\rho(0)\otimes\exp(-\beta H_\text{SB})/Z_\text{SB}$ as
\begin{align}
  \mel{s^+(t)}{\tilde\rho(t)}{s^-(t)} &= \int\mathcal{D}\left[s^\pm(t)\right] e^{i\left(S\left[s^+(t)\right] - S\left[s^-(t)\right]\right)/\hbar}\nonumber\\
  &\times \mel{s^+(0)}{\tilde\rho(0)}{s^-(0)}\mathcal{F}\left[s^\pm(t)\right],
\end{align}
where $s^+(t)$ is the forward path, $s^-(t)$ is the backward path, $S$
is the classical action and $\mathcal{F}$ is the Feynman-Vernon
influence functional~\cite{feynmanTheoryGeneralQuantum1963} that
encodes the system-bath interaction and causes the dynamics to become
non-Markovian in general. The influence functional, $\mathcal{F}$, for
a harmonic bath can be expressed as the exponential of the ``influence phase''
$\exp\left(-\Phi\left[s^\pm(t)\right]\right)$ in terms of the bath
response function, $\alpha(t)$,
\begin{align}
  \Phi\left[s^\pm(t)\right] &= \int_0^t \dd t'\int_0^{t'}\dd t''\Delta s(t') \Re\alpha(t'-t'')\Delta s(t'')\nonumber\\
  &-2i\int_0^t \dd t'\int_0^{t'}\dd t''\Delta s(t')\Im\alpha(t'-t'')\bar{s}(t'')
\end{align}
where $\Delta s(t) = s^+(t) - s^-(t)$ and
$\bar{s}(t) = \frac{1}{2}\left(s^+(t) + s^-(t)\right)$. While these
expressions have been provided here in continuous time, if time is
discretized and a symmetric Trotter splitting is used with the
quasi-adiabatic split Hamiltonian, these expressions become identical
to the ones used in the quasi-adiabatic propagator path integral
method.~\cite{makriTensorPropagatorIterativeI1995,
  makriTensorPropagatorIterativeII1995} If the bath is characterised
by a spectral density, $J(\omega)$, then
\begin{align}
  \alpha(t) &= \frac{1}{\pi}\int_0^\infty \dd \omega J(\omega)\left(\coth\left(\frac{\hbar\omega\beta}{2}\right)\cos(\omega t) - i \sin(\omega t)\right),\label{eq:bath-response}
\end{align}
where $\beta=1/k_BT$ is the inverse temperature for the simulation.

Quantum-classical path
integral\cite{lambertQuantumclassicalPathIntegralI2012,
  lambertQuantumclassicalPathIntegralII2012} (QCPI) by Makri and
coworkers offers a way of estimating the influence functional for
general environments using quasi-classical (Wigner) trajectories. The system's reduced density matrix at a time $t$ is given by
\begin{align}
  &\mel{s^+(t)}{\tilde\rho(t)}{s^-(t)} = \iint \dd q_0 \dd p_0 P(q_0, p_0)\nonumber\\
  &\qquad\qquad\qquad\qquad\times\mel{s^+(t)}{Q_{q_0, p_0}(t)}{s^-(t)},\\
  &\mel{s^+(t)}{Q_{q_0, p_0}(t)}{s^-(t)} = \int\mathcal{D}\left[s^\pm(t)\right]\mel{s^+(0)}{\tilde\rho(0)}{s^-(0)}\nonumber\\
  &\times e^{i\left(S_\text{ref}\left[s^+(t);q_0, p_0\right] - S_\text{ref}\left[s^-(t); q_0, p_0\right]\right)/\hbar} \exp\left(-\Phi_\text{back}\left[s^\pm(t)\right]\right),
\end{align}
where $P$ is the Wigner phase space density and $Q$ is the quantum-influence
function. The reference action, $S_\text{ref}$, is the classical action of the
system along a given path under the time-dependent Hamiltonian, $H_\text{ref}(t;q_0,p_0) = H_0 +
H_\text{SB}(q(t; q_0, p_0))$, generated along the solvent reference trajectory,
$q_\text{ref}(t; q_0, p_0)$, with initial conditions $(q_0, p_0)$. The modified influence
functional is related to the reference-dependent residual back-reaction,
$\Phi_\text{back}$.

QCPI reproduces the exact influence functionals for the case of harmonic baths.
If the reference trajectory is chosen to be the free solvent trajectory, 
\begin{align}
    q(t) &= q_0\cos(\omega t) + \frac{p_0}{\omega}\sin(\omega t),
\end{align}
then it is known that the real part of the bath response function, or the
classical memory is completely accounted for\cite{makriQuantumclassicalPathIntegral2015}. This is called the
ensemble-averaged classical path
reference\cite{banerjeeQuantumClassicalPathIntegral2013}. Consequently,
$\Phi_\text{back}$ becomes the part that simply comes from the imaginary part of
the bath response function or the quantum
memory\cite{makriQuantumclassicalPathIntegral2015}. This can be related to the imaginary part of the $\eta$-coefficients from QuAPI and the bath response function\cite{makriTensorPropagatorIterativeI1995, makriQuantumclassicalPathIntegral2015}.

For a molecular or anharmonic solvent, the general procedure is to solve for $\Phi_\text{back}$ using single classical trajectories for each of the system paths within memory. This exponential proliferation of the classical trajectories can be avoided under the harmonic backreaction approximation\cite{wangQuantumclassicalPathIntegral2019, bosePhaseSpacePath2018} where one uses the anharmonic reference trajectories defined by
\begin{align}
    \dot{q}(t) &= \frac{p(t)}{m}\\
    \dot{p}(t) &= -\nabla V_\text{sol}(q(t))
\end{align}
where $V_\text{sol}$ is the pure solvent environment to get the reference
actions, but replaces the backreaction $\Phi_\text{back}$ with that of the
equivalent harmonic mapping.

While the use of reference propagator significantly decreases the length of the
effective non-Markovian memory and allows for use of larger time-steps, even under harmonic backreaction, the path
integral for the quantum influence function, $Q$, still has to be computed for each Monte Carlo sampled initial
condition, $(q_0, p_0)$. This can often be a major bottleneck especially for
large systems with long ``quantum'' memory.

This motivates the development of an alternative formulation that
eliminates explicit path integral evaluation while retaining the full
non-Markovian memory through a structured representation of the
dissipative kernel. Since the residual backreaction influence functional depends
only on $\Im\alpha(t)$, the quantum influence function $Q_{q_0,p_0}(t)$ has
precisely the same structure as the reduced density matrix of a system would
have except the bath response function is formally replaced by the purely
imaginary $C_\text{back}(t)=i\Im\alpha(t)$. Additionally, the bare system
evolution is now under the trajectory-dependent reference Hamiltonian,
$H_\text{ref}(t; q_0, p_0)$. Consequently, any exponential decomposition of $\Im\alpha(t)$
immediately gives rise to a hierarchy of auxiliary density operators identical
in structure to the standard HEOM. In the present work, we therefore replace the
path-integral evaluation of $Q_{q_0,p_0}(t)$ by propagation of the corresponding
hierarchy and recover the physical reduced density matrix by averaging over the
initial Wigner distribution.

\added{Whenever $C_\text{back}(t)$ admits an exponential representation, either exact or approximate, it can be written as
\begin{align}
    C_\text{back}(t) &= \sum_{m=1}^M c_m \exp(-\nu_m t),\label{eq:exp-decomp}
\end{align}
where $\nu_m\in\mathbb{C}$ are the poles of the exponential representation of
the residual bath response function. In the simplest case of a Drude-Lorentz
spectral density, the poles are all real. In more general cases, that is not
necessary. However, because the bath response function in general, and
$C_\text{back}(t)$ in particular, satisfies $C_\text{back}(-t) =
C^*_\text{back}(t)$, all non-real poles occur in complex conjugate pairs. Rather than treat the real and the complex poles separately, we associate with each exponential mode tuples of three numbers, $(\nu_m,
c_m, \tilde{c}_m)$, where $\nu_m$ and $c_m$ correspond to
Eq.~\eqref{eq:exp-decomp}. The third number $\tilde{c}_m$ is the coefficient
corresponding to $\nu_m$ in the exponential expansion of $C^*_\text{back}(t) =
\sum_m c_m^* \exp(-\nu_m^* t)$. So if $\nu_m\in\mathbb{R}$, the $\tilde{c}_m =
c_m^*$. However, if $\nu_m$ is complex and $\nu_n$ is its complex conjugate,
then $\tilde{c}_m = c_n^*$. Using this, one immediately gets equations for the
quantum influence function, $Q_{q_0, p_0}(t)$, that are formally similar to
standard HEOM. We indicate the auxiliary quantum influence function (AQIF) of
the $\mathbf{n}$th level as $Q^\mathbf{n}_{q_0, p_0}(t)$. Using this notation
now we list the scaled HEOM-like
equations~\cite{shiEfficientHierarchicalLiouville2009} for QC-HEOM}
\begin{align}
    \dot{Q}^\mathbf{n}_{q_0, p_0}(t) &= -\left(i\mathcal{L}_\text{ref}(t; q_0, p_0) + \sum_{m=1}^M n_m \nu_m\right) Q^\mathbf{n}_{q_0, p_0}(t)\nonumber\\
    &- i\left[\hat{s}, \sum_{m=1}^M \sqrt{(n_m+1)s_m} Q^{\mathbf{n}^+_m}_{q_0, p_0}(t)\right]\nonumber\\
    &- i \sum_{m=1}^M\sqrt{\frac{n_m}{s_m}}\left(c_m\hat{s}Q^{\mathbf{n}^-_m}_{q_0, p_0}(t) - \tilde{c}_m Q^{\mathbf{n}^-_m}_{q_0, p_0}(t)\hat{s}\right),\label{eq:qc-heom}
\end{align}
\added{where $s_m = \sqrt{\abs{c_m\tilde{c}_m}}$. For purely real poles, this
equation parallels the scaled HEOM equations of
motion~\cite{shiEfficientHierarchicalLiouville2009}. The exponential
representation may be obtained analytically for several model spectral densities
or numerically using rational approximation techniques such as the barycentric
AAA algorithm~\cite{xuTamingQuantumNoise2022,
nakatsukasaAAAAlgorithmRational2018}. In the latter case, QC-HEOM naturally
admits extensions based on the free-pole HEOM formulation of Xu \textit{et
al.}~\cite{xuTamingQuantumNoise2022} which will be an interesting direction to
pursue in the future.}\comment{QC-HEOM generalized to handle complex exponential representations of the residual response function}

Notice that the low temperature termination is conspicuously absent in
Eq.~\eqref{eq:qc-heom}, because the entire thermal component is taken into
account through the classical trajectories. Here, the modes come solely from the
poles for the spectral density and not from those of the Bose-Einstein
distribution like in case of standard HEOM. $\mathcal{L}_\text{ref}(t; q_0,
p_0)$ is the Liouvillian corresponding to the time-dependent reference
Hamiltonian, $\mathcal{L}_\text{ref}(t; q_0, p_0) = \comm{H_\text{ref}(t; q_0,
p_0)}{\cdot}$. \added{The conceptual framework of QC-HEOM is naturally
compatible with other recent developments of HEOM including incorporation of
tensor networks. Notice that the twin advantage of not requiring any Matsubara
modes, and being able to approximately capture the effect of anharmonic baths
through classical trajectories are independent of the details of the hierarchy
and only dependent on the reference propagators. Consequently, irrespective of
the exact hierarchy and nature of poles, they will continue to hold. These will
be the focus of later developments.}\comment{Discussion on modern tensor network
approaches}

This is similar to the stochastic HEOM
method~\cite{moixHybridStochasticHierarchy2013,
keExtensionStochasticHierarchy2017}, which stochastically unravels the classical
memory by using a colored noise kernel $\xi(t)$. The Hubbard-Stratonovich
transform restricts $\expval{\xi(t)} = 0$, and relates the two-time correlation
function to the real part of the bath response
function\cite{moixHybridStochasticHierarchy2013}. One of the most important
differences here is that instead of an abstract colored noise kernel, QC-HEOM
leverages insights from QCPI to relate the stochastic unraveling to linearised
semiclassical or Wigner trajectories of the free solvent degrees of freedom.
\added{This suggests a possible advantage for future extensions toward systems
beyond the harmonic mapping, where Gaussian statistics and the
Hubbard–Stratonovich transformation may no longer provide an adequate
description.}

For standard HEOM, the decomposition of the bath response function into
exponential decays, Eq.~\eqref{eq:exp-decomp}, gives rise to an infinite set of
Matsubara poles even when the poles of the spectral density are finite in
number. This leads to a convergence problem especially at low temperatures when
accurately representing the Bose-Einstein distribution becomes demanding.
Various techniques like Pad\'e fitting have therefore been developed to reduce
the number of poles required in practical calculations. In QC-HEOM, however, the
hierarchy is constructed only from the residual backreaction kernel
$C_\text{back}(t)=\Im\alpha(t)$. Since $\Im\alpha(t)$ is independent of
temperature, the thermal contribution contained in $\Re\alpha(t)$ is accounted
for entirely through the ensemble of classical reference trajectories.
Consequently, the Matsubara contribution never enters the hierarchy,
\added{trading one of the primary low-temperature bottlenecks of conventional
HEOM in favor of a simulation of multiple trajectories}. Therefore, as will be
demonstrated in Section~\ref{sec:numerics}, in contrast to standard HEOM, the
hierarchy of QC-HEOM is only weakly dependent on the temperature.

\section{Numerical Illustrations}\label{sec:numerics}
We will discuss application of QC-HEOM to the spin-boson model and the
7-site model of the Fenna--Matthews--Olson complex at a variety of temperatures.
Comparisons will be shown with QCPI, QuAPI, and HEOM with Matsubara and Pad\'e
decompositions. \added{In all QC-HEOM and QCPI simulations, a light shaded band
of the same color indicates the standard error associated with the stochastic
sampling of the Wigner trajectories. These uncertainties remain small for all
systems and temperatures considered and are generally smaller than the line
width demonstrating that the number of reference trajectories used provides
statistically converged ensemble averages for the benchmark systems studied
here.}\comment{Comment on trajectory convergence.} All methods were used from
the QuantumDynamics.jl package\cite{boseQuantumDynamicsjlModularApproach2023}.
An open-source implementation of QC-HEOM will be released as a part of the next
update to the framework.

\subsection{Spin-Boson System}
Consider an asymmetric spin-boson system
\begin{align}
    H &= H_0 + \sum_j \frac{p_j^2}{2} + \frac{1}{2}\omega_j^2x_j^2 - c_j x_j \sigma_z,\\
    H_0 &= \sigma_z - \sigma_x,
\end{align}
where $\sigma_{x/z}$ are the Pauli spin matrices. The bath interacts with the
$\sigma_z$ system operator, and is characterized by a Drude spectral density,
\begin{align}
    J(\omega) = \frac{\lambda\gamma\omega}{2(\omega^2+\gamma^2)},
\end{align}
where the reorganization energy $\lambda=1.5$ and the cutoff frequency
$\gamma=7.5$. \added{The simulations are done at an inverse temperature of
$\beta=2.0$.}\comment{$\beta$ added for the Spin-Boson example.}

\begin{figure}
    \centering
    \subfloat[EACP reference QCPI]{\includegraphics{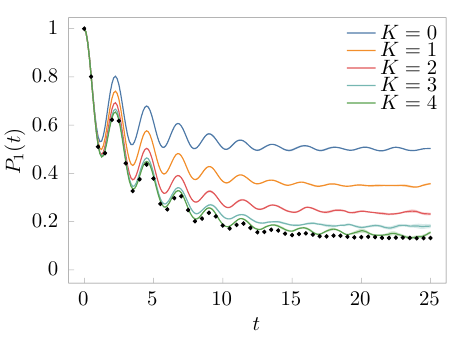}}

    \subfloat[QC-HEOM]{\includegraphics{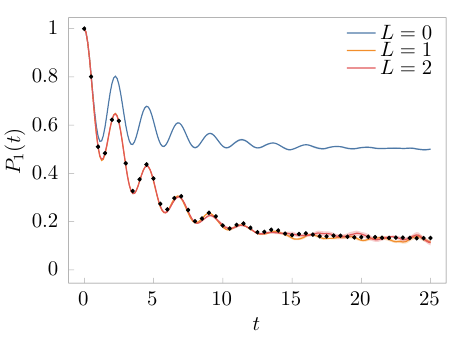}}
    \caption{Comparison of convergence of EACP-reference QCPI and QC-HEOM for the spin-boson model. Black markers: Converged QuAPI results.}
    \label{fig:spin-boson}
\end{figure}

We demonstrate the route of convergence of QC-HEOM to the QuAPI
results and compare it with QCPI results with a free solvent (or EACP)
reference. The results are shown in Fig.~\ref{fig:spin-boson}. First,
note that QC-HEOM reproduces the EACP results at $L=0$. This
corresponds to the $K=0$ EACP run for QCPI. QCPI shows a gradual
convergence to the QuAPI results (shown in black markers) with memory
length $K$. It is well-known that using a dynamically consistent
state-hopping~\cite{waltersIterativeQuantumclassicalPath2016} (DCSH)
reference can speed up convergence, capturing effects of the asymmetry
even at $K=0$ levels. Coming to the QC-HEOM convergence, though it is
controlled by the depth of hierarchy and not the memory length, we see
certain qualitatively different features. Increasing the hierarchy
depth from $L=0$ to $L=1$ remarkably is already sufficient to obtain
near-quantitative agreement with the QuAPI benchmark.

At $L=0$, QC-HEOM solves the equation of motion for a single $2\times 2$ AQIF.
Because there is only a single Drude bath, the imaginary part of the bath
response function can be represented by a single exponential. Therefore, the
$L=1$ calculation only needs to solve coupled differential equations involving
two $2\times 2$ AQIFs. Finally, at $L=2$, the differential equations involve
three AQIFs. The rapid convergence with hierarchy depth reflects a fundamental
difference between the two approaches. QCPI approximates the bath memory by
truncating the influence functional after $K$ timesteps, whereas QC-HEOM encodes
the bath memory through the hierarchy of auxiliary quantum influence
functionals. Increasing $L$ systematically improves the representation of
non-Markovian effects without introducing an explicit memory cutoff.

\begin{figure}
    \centering
    \subfloat[Composite spectral density]{\includegraphics{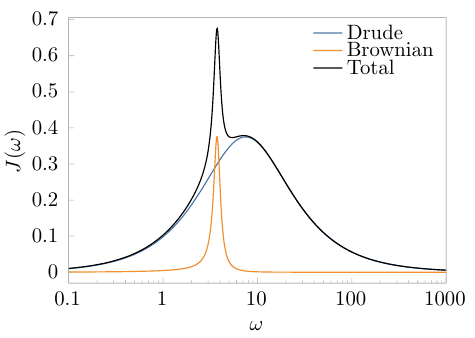}}

    \subfloat[Dynamics]{\includegraphics{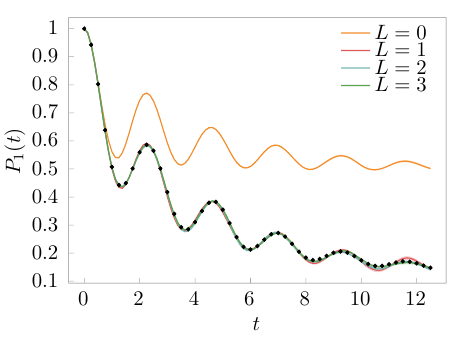}}
    \caption{Composite spectral density and the corresponding dynamics. Black markers are converged path integral results with 60 steps of memory.}
    \label{fig:composite-dyn1}
\end{figure}
\added{As an example with a more involved spectral density, consider a spectral density that is given as a sum of a Drude-Lorentz and an underdamped Brownian oscillator:
\begin{align}
    J(\omega) = \frac{\lambda_D\gamma_D\omega}{2(\omega^2+\gamma_D^2)} + \frac{\lambda_U\gamma_U\omega_0^2\omega}{2((\omega^2-\omega_0^2)^2+\gamma_U^2\omega^2)},
\end{align}
where $\lambda_D$ and $\lambda_U$ are the reorganization energies corresponding
to the Drude-Lorentz and the underdamped Brownian baths respectively, $\gamma_D$
and $\gamma_U$ are their respective time-scales, and $\omega_0$ is the center of
the underdamped oscillator. The combined bath interacts with the $\sigma_z$
operator of the system. One can think of this as modeling an unstructured bath
with a single important frequency. For the first example, we simulated
$\lambda_D=10\lambda_U=1.5$ yielding a total reorganization energy of 1.65. The
timescales were chosen to be $\gamma_D=7.5$ and $\gamma_U=0.75$, and
$\omega_0=3.75$. The resultant spectral density and the dynamics are shown in
Fig.~\ref{fig:composite-dyn1}. We once again see the characteristic convergence
of QC-HEOM at very low depths of the hierarchy. A value of $L=2$ is completely
converged, with $L=1$ capturing all the physically relevant properties but
slightly overestimating the long time oscillations. In contrast the path
integral simulations required extremely long memory lengths owing to the
underdamped Brownian oscillator. The time-evolving matrix product operators
algorithm~\cite{strathearnEfficientNonMarkovianQuantum2018} was used and
convergence was achieved at a memory length of 60 time-steps.}

\subsection{Fenna--Matthews--Olson Complex}
\added{The Fenna--Matthews--Olson complex is one of the most widely studied complex involved in photosynthetic exciton transport.} We demonstrate the method using the 7-site Frenkel-Holstein model of the
Fenna--Matthews--Olson complex developed and studied by Ishizaki and
Fleming\cite{ishizakiTheoreticalExaminationQuantum2009},
\begin{align}
    H &= H_0 + \sum_{b=1}^7\sum_j \frac{p_{bj}^2}{2} + \frac{1}{2}\omega_{bj}^2 x_{bj}^2 - c_{bj} x_{bj}\dyad{b},\\
    H_0 &= \sum_{s=1}^7 \epsilon_s \dyad{s} + \sum_{j=1}^7\sum_{k=j+1}^7 h_{jk}\left(\dyad{j}{k} + \dyad{k}{j}\right),
\end{align}
where $\epsilon_s$ is the Frank-Condon excitation energy of the $s$th site, and
$h_{jk}$ is the electronic coupling between the $j$th and the $k$th sites.
Every site of this excitonic model is coupled with independent harmonic baths
described by identical Drude-Lorentz spectral densities,
\begin{align}
    J_j(\omega) &= \frac{2\lambda_j\gamma_j\omega}{\omega^2+\gamma_j^2},
\end{align}
where $\lambda_j=\SI{35}{\per\cm}$ is the reorganization energy of the $j$th
bath and characteristic phonon frequency of $\gamma_j=\SI{50}{\per\cm}$ is
taken~\cite{ishizakiTheoreticalExaminationQuantum2009}. \added{Since the
original publication of these parameters, several thorough experimental and
theoretical studies have revealed that the vibronic reorganization energy used
in this model is probably unrealistically
small~\cite{duanQuantumCoherentEnergy2022}, leading to the simulations showing
long-lasting coherent oscillations. Several alternate spectral densities have
been suggested over the
years~\cite{riveraInfluenceSiteDependentPigment2013,maityDFTBMMMolecular2020}.
Using the parameters of Maity, \textit{et. al}~\cite{maityDFTBMMMolecular2020},
we have simulated the dynamics using path integrals and explored the transport
mechanism in the FMO complex~\cite{boseImpactSolventStatetoState2023,
boseImpactSpatialInhomogeneity2023, boseIncorporationEmpiricalGain2024}.
}\comment{Discussion of recent spectral densities and the low reorganization energy of the Ishizaki-Fleming model.}

\begin{figure*}
    \subfloat[$T=\SI{300}{\kelvin}$]{\includegraphics{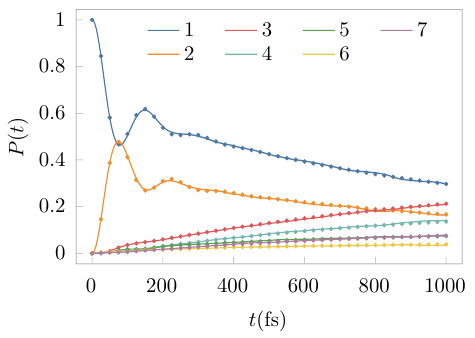}}
    ~\subfloat[$T=\SI{77}{\kelvin}$]{\includegraphics{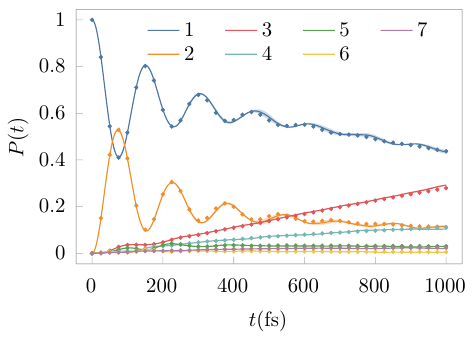}}
    
    \subfloat[$T=\SI{10}{\kelvin}$]{\includegraphics{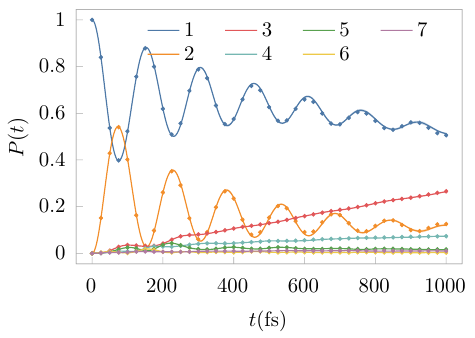}}
    ~\subfloat[Convergence pattern]{\includegraphics{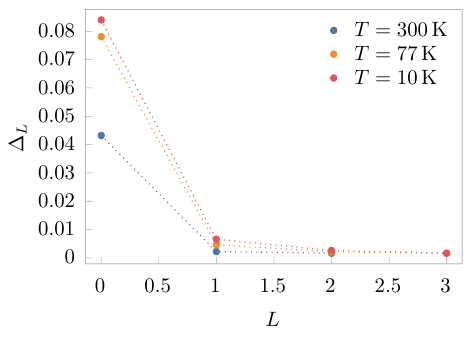}}

    \caption{Population dynamics in the 7-site Fenna--Matthews--Olson complex\cite{ishizakiTheoreticalExaminationQuantum2009} at different temperatures. Markers show the standard HEOM simulation results as reference. Lines denote the QC-HEOM results.}\label{fig:fmo}
\end{figure*}

The backreaction for the $s$th site is governed by
\begin{align}
    C^{s}_\text{back}(t) &= -\frac{i}{\pi}\int_0^\infty \dd\omega J_s(\omega)\sin(\omega t)\\
    &= -i\lambda_s\gamma_s\exp(-\gamma_s t).
\end{align}
Since $C^s_\text{back}(t)$ depends only on the imaginary part of the bath response
function, each Drude-Lorentz bath contributes exactly one exponential term
corresponding to the physical Drude pole. In contrast to conventional HEOM, no
Matsubara or Padé poles associated with the thermal factor are required.
Consequently, the hierarchy is indexed by a seven-dimensional multi-index
$\mathbf n=(n_1,\ldots,n_7)$, with one hierarchy coordinate associated with each
bath.

Suppose we truncate the hierarchy at a depth of $L$. For a system
coupled to $N_\text{bath}$ independent baths, each contributing a
single hierarchy coordinate, this retains all AQIFs satisfying
$\sum_{j=1}^{N_\text{bath}} n_j \le L$. The total number of AQIFs is
therefore
$\sum_{\ell=0}^{L}
\binom{\ell+N_\text{bath}-1}{\ell}=\binom{L+N_\text{bath}}{L}$. For
the seven-site FMO model ($N_{\text{bath}}=7$), this yields eight
AQIFs at $L=1$ and 36 AQIFs at $L=2$. In contrast to conventional
HEOM, the number of poles is determined solely by the number of
physical baths and is independent of the number of Matsubara or Pad\'e
poles used to represent thermal fluctuations.

\begin{table*}
\caption{Comparison of converged standard HEOM and QC-HEOM calculations for the
FMO complex. Illustrative timing comparisons provided using an adaptive
Runge-Kutta algorithm with coefficients by
Tsitouras~\cite{tsitourasRungeKuttaPairs2011}.}
\label{tab:fmo_scaling}
\begin{tabular}{cccccccccc}
\toprule
$T$ (K) & Decomposition & $N_\text{modes}$ & $L$ & $N_\text{ADO}$ & $L_\text{QC-HEOM}$ & $N_\text{AQIF}$ & $N_\text{ADO}/N_\text{AQIF}$ & HEOM Timing & QC-HEOM Timing\\
\midrule
300 & Matsubara & 0 & 4 & 330 & 1 & 8  & 41.3 & \SI{148}{\sec} & \SI{54}{\sec} / initial condition\\
77  & Matsubara & 3 & 3 & 4495 & 2  & 36 & 155.0 & \SI{2880}{\sec} & \SI{240}{\sec} / initial condition\\
10  & Pad\'e & 6 & 3 & 22100 & 2 & 36 & 613.9 & \SI{15595}{\sec} & \SI{250}{\sec} / initial condition\\
\bottomrule
\end{tabular}
\end{table*}

We study the system at three distinct temperatures: (a) physiological
temperature of \SI{300}{\kelvin}, (b) liquid nitrogen temperature of
\SI{77}{\kelvin}, and (c) \SI{10}{\kelvin}. For the QC-HEOM calculation, each of
the site-local Drude-Lorentz baths are discretized using 50 oscillators. A total
of 10000 independent reference classical trajectories are sampled from the
thermal Wigner distribution. We compare the results for the population dynamics
with converged standard HEOM results. This is shown in
Figs.~\ref{fig:fmo}~(a)--(c). To quantify the convergence of QC-HEOM toward the
benchmark results, we define the time-averaged root-mean-square (RMS) population
error for site j as \begin{align} \Delta_j &=
\sqrt{\frac{1}{t_\text{fin}}\int_0^{t_\text{fin}}
\left(\mel{j}{\tilde{\rho}_\text{HEOM}}{j} -
\mel{j}{\tilde{\rho}_\text{QC-HEOM}}{j}\right)^2}, \end{align} where
$t_\text{fin}=\SI{1}{\ps}$ is the final time of the simulation done. The total
per-site error measure $\Delta$ is defined as the average of the RMS errors over
all sites. We demonstrate the behavior of this error for all three temperatures
on increasing the depth of hierarchy in Fig.~\ref{fig:fmo}~(d).

At $T=\SI{300}{\kelvin}$, the system is in the high temperature regime and
Matsubara poles are not even required for the benchmark calculation. Convergence
is reached at a hierarchy depth of $L=4$ using 330 auxiliary density operators
(ADOs). For QC-HEOM, we see that at $L=1$, we already have agreement with the
HEOM results within numerical errors. The coupled differential equations in this
case use 8 AQIFs. The excellent agreement demonstrates that even at high
temperatures, the inclusion of reference trajectories leads to convergence of
dynamics at shallower hierarchies.

Next coming to the liquid nitrogen temperature of $T=\SI{77}{\kelvin}$, the
benchmark calculations require three Matsubara poles for convergence of the
Bose-Einstein distribution. A hierarchy depth of $L=3$ is required, amounting to
a total of 4495 ADOs. In contrast, the QC-HEOM as expected still does not
require any poles for description of the thermal distribution. The residual
quantum memory is accurately captured already at the level of $L=2$. Thermal
fluctuations do not contribute to the hierarchy and instead enter through the
ensemble average over trajectories sampled from the thermal Wigner distribution.
Consequently, lowering the temperature increases the hierarchy depth only mildly
despite the substantial increase in the number of Matsubara terms required by
the benchmark HEOM calculation.

At both of these temperatures, much like the spin-boson, the majority of
correction happens from $L=0$ to $L=1$ level. Further corrections happen at
$L=2$, where even at this low temperature, we get quantitative agreement with
HEOM results. The $L=0$ calculations use a single AQIF as expected. However, in
this case, because of the presence of multiple baths, $L=1$ calculations
propagate coupled differential equations involving eight $7\times 7$ AQIFs.

To push the limits of QC-HEOM, we also simulated the system at
$T=\SI{10}{\kelvin}$. At this low temperature, the number of Matsubara
modes required to converge the Bose-Einstein distribution was
extremely large.  Therefore, we used a Pad\'e-based approach for
expressing the bath response function as a sum-over-poles for the
standard HEOM\cite{huCommunicationPadeSpectrum2010,
  huPadeSpectrumDecompositions2011}. With the Pad\'e decomposed
approach to HEOM, six poles were required to converge the thermal
component.  Additionally, a total depth of three is required for
convergence of dynamics, leading to a calculation with 22100
ADOs. Once again, in contrast, we see that QC-HEOM is qualitatively
correct at $L=1$ and finally converges by $L=2$ using 36 AQIFs.

\added{The statistics corresponding to all of these runs is summarized in
Table~\ref{tab:fmo_scaling}. While the number of ADOs or AQIFs is directly
proportional to their individual computational complexities and runtime, because
the base differential equations are not the same, it is not simple to
extrapolate a reduction in these numbers to a speed up. As an illustration of
the translation of this reduction to the computational complexity, we have also
added runtime for the HEOM calculation. The timings for QC-HEOM has been added
in a ``per-trajectory'' manner because the ensemble is embarassingly parallel
and can be distributed to multiple processors and threads independently. All the
differential equations are solved using the adaptive Runge-Kutta parameterized
by Tsitouras~\cite{tsitourasRungeKuttaPairs2011} with an absolute tolerance of
$10^{-6}$. While several better alternatives exist in propagating the equations
of motion, this gives a ballpark estimate, illustrating the reduction in
complexity brought in by the quantum-classical formalism. Future work will focus
on incorporating ideas of tensor network propagation and other exponential
schemes.}\comment{Runtime statistics added.}

\begin{figure}
    \centering
    \includegraphics{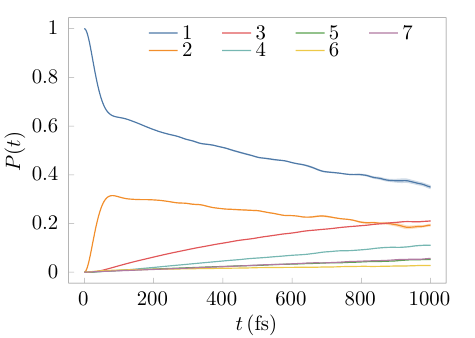}
    \caption{Population dynamics of the FMO system with $\lambda_j = \SI{150}{\per\cm}$ at $\SI{300}{\kelvin}$.}
    \label{fig:FMO150}
\end{figure}

\added{As mentioned earlier, the reorganization energy of the vibrational environments that are closer to current studies is much higher than the models used here. To examine the performance in a more strongly coupled environmental regime, we repeated the simulations using an alternative spectral density with increased reorganization energy, $\lambda_j = \SI{150}{\per\cm}$. The number of oscillators per bath was still taken to be 50, but $10^5$ Monte Carlo points were required to decrease the stochastic variance. Figure~\ref{fig:FMO150} shows the population dynamics for this system. Convergence was reached with $L=3$ in this case. One sees that higher depths of the hierarchy are required as the reorganization energy increases. However, this increase is still better controlled than in conventional HEOM.}\comment{Higher reorganization energy FMO}

\section{Conclusion}\label{sec:conclusion}\comment{The conclusion has been rewritten}
Quantum-classical path integrals provide a rigorous framework for
approximating non-Markovian quantum dynamics in complex environments by
separating the influence of classical solvent fluctuations from the residual
quantum memory. In this framework, the ensemble-averaged classical path
(EACP) reference propagators incorporate the contribution of the real part of
the bath response function through solvent-driven trajectories, while the
remaining quantum influence function retains the non-classical backreaction.
However, despite the reduction in effective quantum memory, QCPI still requires
the evaluation of a path integral for each sampled solvent trajectory, which
can become computationally prohibitive for large systems.

In this work, we have introduced quantum-classical hierarchical equations of
motion (QC-HEOM), which combines the reference trajectory formulation of QCPI
with a hierarchical representation of the residual quantum influence
functional. By explicitly separating the classical and quantum contributions
to the bath response, QC-HEOM replaces the repeated path-integral evaluation
of the quantum influence function by a hierarchy of coupled auxiliary quantum
influence functionals. For harmonic environments, this construction provides a
formally exact reformulation of the residual quantum memory and is equivalent
to the underlying QuAPI description.

A central consequence of this separation is that the hierarchy is constructed
only from the temperature-independent residual backreaction kernel. Unlike
conventional HEOM, where the thermal component of the bath correlation
function requires a Matsubara or Pad\'e decomposition, QC-HEOM incorporates
thermal fluctuations entirely through the ensemble of Wigner trajectories.
Consequently, the hierarchy size exhibits only weak temperature dependence and
is determined primarily by the physical poles of the spectral density rather
than by the poles required to represent the Bose-Einstein distribution. This
removes one of the primary sources of low-temperature complexity in standard
HEOM.

The performance of QC-HEOM was demonstrated for an asymmetric spin-boson model
and the seven-site Fenna--Matthews--Olson complex over a broad range of
temperatures. In both cases, quantitative agreement with benchmark QuAPI and
HEOM calculations was obtained using remarkably shallow hierarchy depths. For
the FMO complex, QC-HEOM required only 8--36 auxiliary quantum influence
functionals, compared with hundreds to tens of thousands of auxiliary density
operators in conventional HEOM calculations. At $T=10$ K, for example, the
Pad\'e-decomposed HEOM calculation required 22100 auxiliary density operators,
whereas QC-HEOM converged with 36 auxiliary quantum influence functionals,
corresponding to a reduction of approximately a factor of 614 in the number of
hierarchical objects. While this reduction directly reflects the reduced
hierarchical complexity, the overall computational advantage depends on the
implementation details, trajectory sampling strategy, and numerical
propagation scheme.

An important feature of QC-HEOM is that the hierarchy is not intrinsically
limited to a particular choice of reference dynamics. The present work employs
the EACP reference because it provides an exact separation of classical and
quantum memory for harmonic baths. However, more physically motivated reference
trajectories may further reduce the residual quantum contribution that must be
represented hierarchically. In particular, exploring trajectories generated from
mean-field approaches such as Ehrenfest dynamics, or the dynamically consistent
state-hopping reference~\cite{waltersIterativeQuantumclassicalPath2016} seem to
be promising choices. Such optimized reference propagators could shift
additional environmental effects from the quantum hierarchy into the classical
trajectory ensemble, thereby reducing the required hierarchy depth while
retaining a systematic treatment of residual quantum memory.

The present formulation focuses on harmonic environments, where the QC-HEOM
construction is formally exact. Nevertheless, the trajectory-based nature of
the method provides a natural route toward extensions involving anharmonic and
atomistic environments. Similar to QCPI, reference trajectories generated from
molecular simulations or \textit{ab initio} dynamics could be incorporated,
while the remaining quantum backreaction could be treated using suitable
approximations such as harmonic mappings or improved semiclassical corrections
\cite{wangQuantumclassicalPathIntegral2019,
bosePhaseSpacePath2018}. The development of physically optimized reference
trajectories and systematic approximations for anharmonic residual memory
represent important directions for future work.

Finally, the hierarchical structure introduced here is complementary to recent
developments in HEOM methodology, including generalized exponential decompositions and
free-pole representations~\cite{xuTamingQuantumNoise2022}, tensor-network
representations of the hierarchy\cite{chenTreeTensorNetwork2025}, and other approaches for reducing the
complexity of large auxiliary spaces. Combining these developments with the
quantum-classical separation introduced in QC-HEOM may enable simulations of
larger molecular and condensed-phase systems with structured and anharmonic
environments.

Overall, QC-HEOM provides a new perspective on hierarchical open-system
dynamics by treating classical environmental fluctuations through trajectory
ensembles while reserving the hierarchy for the genuinely quantum component of
the environmental memory. This separation offers a promising framework for
combining the accuracy of numerically exact methods with the flexibility of
trajectory-based approaches for increasingly complex quantum dynamical
systems.

\appendix
\section{Exponential representation of the temperature-independent bath kernel}\label{sec:exp-ref-cback}\comment{Appendix discussion exponential decomposition}
For a general spectral density, $J(\omega)$, the bath response function, $\alpha(t)$ given by Eq.~\eqref{eq:bath-response}, governs HEOM, which proceeds to decompose it in terms of exponentials. In QC-HEOM, $\Re \alpha(t)$ is fully incorporated through the classical trajectories
initialized on the thermal Wigner distribution for the bath. It then requires
exponential decompositions of the $C_\text{back}(t)$. While the current
implementation uses a more traditional HEOM-like backend, the concept of QC-HEOM, by itself, is not limited to such a
treatment. As mentioned in the main body of the paper, ideas of barycentric
representations using the AAA
algorithm~\cite{nakatsukasaAAAAlgorithmRational2018} or other rational fittings
for more general spectral densities, as employed by the free-pole
HEOM~\cite{xuTamingQuantumNoise2022} can be adapted for $C_\text{back}(t)$ in
the future. This will allow QC-HEOM to be systematically extended to an even broader
class of spectral densities.

In terms of model spectral densities, we have already mentioned that for the
Ohmic form with a Lorentzian tail, conventionally called the Drude-Lorentz
spectral density, $C_\text{back}(t)$ is a single exponential with real poles.
Here, we discuss the decompositions of $C_\text{back}(t)$ for some common types
of model spectral densities. For an Ohmic spectral density with a Lorentzian
decay, or what is traditionally called a Drude-Lorentz spectral density, one can
show that $C_\text{back}(t)$ can be represented as a single exponential decay. 
\begin{align}
    J^\text{ohmic}(\omega) &= \frac{2\lambda\gamma}{\Delta s^2}\frac{\omega}{\omega^2+\gamma^2}\\
    C^\text{ohmic}_\text{back}(t) &= -i\frac{\lambda\gamma}{\Delta s^2}\exp(-\gamma t)
\end{align}

Now consider a more general family of spectral densities, all with the Lorentzian cutoff,
\begin{align}
    J^{(s)}(\omega) &= \frac{2\lambda\gamma^{2-s}}{\Delta s^2} \sin\left(\frac{\pi s}{2}\right)\frac{\omega^s}{\omega^2 + \gamma^2},
\end{align}
for $s>0$, where $s=1$ reduces to the familiar Ohmic case, and $0<s<1$ are their
sub-Ohmic counterparts. The set of spectral densities generated by using $s>1$
seems to generate a family of super-Ohmic spectral densities. The
Lorentzian-cutoff family generated here is not suitable for physically
meaningful super-Ohmic spectral densities because its reorganization energy
diverges for $s>1$. Realistic super-Ohmic environments instead employ rapidly
decaying high-frequency cutoffs, such as Gaussian or exponential, or
experimentally parameterized spectral densities.

In contrast to the Ohmic Drude-Lorentz, these sub-Ohmic cases do not lead to
simplification in the sense that the bath response function can be represented
by a single pole. There are two contributions to $C_\text{back}(t)$, one from
the pole and the other from a branch cut. This latter part is not exponential
but polynomial in $t$:
\begin{align}
    C^{s}_\text{back}(t) &= -i\frac{\lambda\gamma}{\Delta s^2}\sin^2\left(\frac{\pi s}{2}\right)\exp(-\gamma t)\nonumber\\
    &- i\frac{2\lambda\gamma^{2-s}}{\pi\Delta s^2}\Gamma(s+1)\sin\left(\frac{\pi s}{2}\right)\sin(\pi s)t^{-(s+1)}.
\end{align}
The first term is an exponential decay and fits our requirement. Now for
QC-HEOM, the second term needs to be expanded in terms of exponentials. Laplace
transforming the algebraic tail, we get
\begin{align}
    t^{-(s+1)} &= \frac{1}{\Gamma(s+1)}\int_0^\infty \Lambda^s \exp(-\Lambda t) \dd\Lambda.
\end{align}
So one can in principle simply discretize the integral into several modes through a quadrature representation. However, a na\"ive linear grid quadrature would involve a proliferation of the number of modes. A more optimal approach is to use a logarithmic discretization by assuming $\Lambda=\exp(x)$ and consequently $\dd\Lambda = \exp(x)\dd x$. Now, $x$ can be uniformly discretized, resulting in a geometric spacing of the $\Lambda$ points. The final controlled exponential decomposition of the algebraic tail using this discretization, therefore, is:
\begin{align}
    x_k &= x_\text{min} + k\Delta x\\
    \Lambda_k &= \exp(x_k)\\
    t^{-(s+1)} &= \sum_{k=1}^N \frac{\Delta x}{\Gamma(s+1)}\Lambda_k^{s+1}\exp(-\Lambda_kt)
\end{align}
Thus, although the sub-Ohmic kernel is not represented exactly by a finite
number of exponential terms, it admits a systematically improvable exponential
expansion through quadrature of the Laplace representation.

For the Brownian baths, where
\begin{align}
    J(\omega) &= \frac{2\lambda}{\Delta s^2}\frac{\gamma\omega_0^2\omega}{\left(\omega^2-\omega_0^2\right)^2 + \gamma^2\omega^2},
\end{align}
the residual bath response function is expanded as a sum over two exponentials. If we define $\Omega = \sqrt{\omega_0^2 - \frac{\gamma^2}{4}}$, then the residual bath response function,
\begin{align}
    C_\text{back}(t) &= -\frac{\lambda \omega_0^2}{2\Delta s^2\Omega}\left(e^{-\left(\frac{\gamma}{2}-i\Omega\right)t} - e^{-\left(\frac{\gamma}{2}+i\Omega\right)t}\right).
\end{align}
This holds so long as $\Omega\ne 0$, with $\Omega\in\mathbb{R}$ characterizing
the underdamped cases, and purely imaginary $\Omega$s characterizing overdamped
cases. For the underdamped Brownian bath, the poles are conjugates of each
other, whereas they just become two distinct real exponential decays for the
overdamped case. If $\gamma = 2\omega_0$, there is a double pole, which does not
have a simple exponential representation. This critical damping case therefore
requires a slight modification of the hierarchy analogous to the treatment of
repeated poles in generalized HEOM formulations.

\bibliography{references}
\end{document}